\documentclass[aip,jcp,reprint]{revtex4-2}
\usepackage{amsmath,amssymb}
\usepackage{graphicx}
\usepackage{braket}
\usepackage{algorithm}
\usepackage{algpseudocode}
\usepackage{hyperref}
\hypersetup{colorlinks,linkcolor=blue,anchorcolor=blue,citecolor=blue,urlcolor=blue}
\usepackage{orcidlink}
\usepackage{xcolor}

\begin{document}

\title{Bootstrap Embedding for Interacting Electrons in Phonon Coherent-state Mean Field}

\author{Shariful Islam\orcidlink{0000-0001-5230-6499}}
\affiliation{Department of Physics, North Carolina State University, Raleigh, NC 27695, USA}

\author{Joel Bierman}
\affiliation{Department of Electrical and Computer Engineering, North Carolina State University, Raleigh, NC 27695, USA}

\author{Yuan Liu\,\orcidlink{0000-0003-1468-942X}}
\email{q\_yuanliu@ncsu.edu}
\affiliation{Department of Electrical and Computer Engineering, North Carolina State University, Raleigh, NC 27695, USA}
\affiliation{Department of Computer Science, North Carolina State University, Raleigh, NC 27695, USA}
\affiliation{Department of Physics, North Carolina State University, Raleigh, NC 27695, USA}


\begin{abstract}
We develop a fermi-bose bootstrap embedding (fb-BE) framework for the ground state of interacting electrons coupled to phonon mean field. The method combines bootstrap embedding for correlated electrons with a self-consistent coherent-state mean-field treatment for phonons. This method models the interacting electron-phonon problem as a system of correlated electrons traveling in a self-consistently specified potential landscape, allowing for efficient treatment of large lattice systems. Convergence of the methods for fragment size and total system size are demonstrated for one-dimensional Hubbard-Holstein model for up to 350 sites. Finite-size scaling is performed to extrapolate to infinite system size. Benchmarking against density matrix renormalization group for small 8-site system at half- and quarter-filling shows orders-of-magnitude runtime advantage. The comparison further reveals that  the method performs best in regimes dominated by localization, such as the Mott insulating phase and the strong-coupling tiny polaron regime, where the local embedding ansatz is still valid. However, due to the mean-field treatment for phonons, we find limitations of our methods in the weakly coupled delocalized region and at the Peierls transition, where quantum phonon fluctuations and long-range kinetic correlations become substantial. 
\end{abstract}

\maketitle

\section{Introduction}
Strongly correlated electron-phonon systems play a key role in condensed matter physics, including polaron formation, charge-density-wave (CDW) order, metal-insulator transitions, and unconventional superconductivity \cite{Morosan2012,Dagotto2008, Giustino2017}. A quantitative description of these systems is notoriously difficult due to the competing roles of electron-electron and electron-phonon interactions, as well as the exponential expansion of the many-body Hilbert space with system size.

The Hubbard--Holstein model combines local Coulomb repulsion with site-resolved coupling to local phonon modes \cite{Holstein1959, Hubbard1963,Hardikar2007}, creating a simple yet physically rich framework for exploring this interplay. Despite its apparent simplicity, the model reveals a complicated phase diagram driven by the competition between electronic correlations and lattice distortions \cite{Hardikar2007, Clay2005}.

There are many numerical methods developed to study the Hubbard–Holstein model. Exact diagonalization (ED) or full configuration interaction (FCI) give numerically exact results, but can only be used on small clusters because the Hilbert space grows exponentially \cite{Wang2020, Dobry1995}. The density matrix renormalization group (DMRG) method has high accuracy in one dimension \cite{Tezuka2005, Tezuka2007, Jansen2020}, but its computational cost increases dramatically in higher dimensions or for systems with long-range entanglement. Quantum Monte Carlo (QMC) approaches scale more favorably, but suffer from the fermion sign problem or phase problem for bosons \cite{Luo2025,Foulkes2001,Purwanto2004,Rubenstein2012}. Dynamical mean-field theory (DMFT) accurately captures local temporal correlations but ignores non-local spatial fluctuations necessary for low-dimensional ordering, unless computationally expensive cluster extensions are used \cite{Werner2007, Jeon2004, Backes2023}.

For numerical simulation of electron-phonon coupled systems, the primary bottleneck of using these approaches is the simultaneous consideration of the continuous bosonic degrees of freedom and the fermionic correlation problem. The phonon Hilbert space is infinite, therefore truncations may cause artifacts in the strong-coupling regime. By leveraging quantum hardware, these challenges can be partially alleviated \cite{McArdle2020,Macridin2018, Sawaya2020,Bidart2025arxiv,Liu2024arxiv,Crane2024arxiv2409,Kemper2025arxiv2511}. However, there is still a critical need for efficient frameworks that can bridge the gap between small systems and thermodynamic limit.

Embedding methods \cite{Jones2020JACS,Rossmannek2023JPCL,SunChan2016AccChemRes} provide a powerful solution to treat large problems on small computers by separating highly correlated sections (fragments) and treating their interactions with the environment in a self-consistent manner. Density matrix embedding theory (DMET) \cite{Knizia2012, Sandhoefer2016} and bootstrap embedding (BE) \cite{Ye2019JCTC, Meitei2023PeriodicBE,Liu2023JCTC,Bierman2025ChemRxiv} provide high accuracy for exclusively fermionic systems. Bootstrap embedding is especially useful since it treats the problem as a constraint optimization problem with the constraint as matching conditions on reduced density matrices (RDMs) on overlapping regions of adjacent fragments, rather than solving for an explicit bath wavefunction. This enables flexible and systematic matching of local observables between overlapping chunks. 

In this paper, we extend bootstrap embedding to electron-phonon systems by incorporating a mixed fermi-bose eigensolver that treat electrons at a correlated level while treating phonons as a coherent-state mean field (CSMF) \cite{Reinhard2019,Sandhoefer2016}. 
By expressing phonons as site-local coherent displacements, the bosonic sector is essentially turned into a self-consistent potential that acts on electrons. This avoids explicit truncation of the phonon number while capturing the fundamental mechanics of static lattice deformation.
Our technique, unlike quantum implementations, works on classical hardware. The BE matching conditions handle electronic correlations, while the CSMF ansatz transforms the bosonic sector into a self-consistent potential. 

Convergence of the methods for fragment size and total system size are demonstrated for one-dimensional Hubbard-Holstein model for up to 350 sites. Finite-size scaling is performed to extrapolate to infinite system size. Benchmarking against density matrix renormalization group for small 8-site system at half- and quarter-filling suggests orders-of-magnitude runtime advantage, despite limitations for cases where quantum fluctuations of phonon dominate. This solution allows for scalable simulations of the Hubbard-Holstein model beyond the reach of accurate diagonalization while preserving controllable precision in both the electronic and phononic sectors.

The remaining sections of this work are organized as follows. Section~\ref{sec:method} introduces the Hubbard-Holstein Hamiltonian, the coherent-state description of phonons, the fb-BE formalism, and the computational algorithm. Section~\ref{sec:comp} discusses the computational details. Section~\ref{sec:results} presents benchmarking results in different parameter regimes. Section~\ref{sec:conclu} concludes the paper.

\section{Methodology}
\label{sec:method}

We present details on the methodology in this section. Sec.~\ref{ssec:hubbard-holstein} reviews the basics of the Hubbard-Holstein Hamiltonian of interest. Sec.~\ref{ssec:mf-decoupling} presents the coherent-state mean-field approach used in our eigensolver. Sec.~\ref{ssec:fbbe-alg} presents details of the fb-BE algorithm.

\subsection{Hubbard--Holstein Hamiltonian}
\label{ssec:hubbard-holstein}
The Hubbard–Holstein model provides a widely used theoretical framework for studying the interplay between strong electron–electron interactions and local electron–phonon coupling in correlated materials, especially low-dimensional systems such as organic conductors, quasi-1D charge-transfer salts, and transition-metal oxides.\cite{Hubbard1963,Holstein1959,Berger1995}

The total Hamiltonian $\hat{H}$ comprises three parts:
\begin{equation}
\hat{H} = \hat{H}_{\text{el}} + \hat{H}_{\text{ph}} + \hat{H}_{\text{el-ph}},
\end{equation}
where $\hat{H}_{\text{el}}$ represents the purely electronic degrees of freedom with on-site Hubbard interactions, $\hat{H}_{\text{ph}}$ encompasses local phonons at each site, and $\hat{H}_{\text{el-ph}}$ characterizes the Holstein-type electron-phonon coupling.

\paragraph*{Electronic Hamiltonian.} The electronic part of the Hamiltonian is given by:
\begin{equation}
\hat{H}_{\text{el}} = -t \sum_{\langle i, j \rangle, \sigma} \left( \hat{c}_{i\sigma}^\dagger \hat{c}_{j\sigma} + \text{H.c.} \right)
+ U \sum_i \hat{n}_{i\uparrow} \hat{n}_{i\downarrow},
\label{eq:HH_el}
\end{equation}
where $\hat{c}_{i\sigma}^\dagger$ ($\hat{c}_{i\sigma}$) creates (annihilates) an electron of spin $\sigma$ at site $i$, $\hat{n}_{i\sigma} = \hat{c}_{i\sigma}^\dagger \hat{c}_{i\sigma}$ is the number operator, $t$ is the nearest-neighbor hopping amplitude, and $U$ is the strength of the on-site Coulomb repulsion.

\paragraph*{Phonon Hamiltonian.} The phononic degrees of freedom are represented as independent local harmonic oscillators:
\begin{equation}
\hat{H}_{\text{ph}} = \omega_0 \sum_i  \hat{b}_i^\dagger \hat{b}_i ,
\label{eq:HH_ph}
\end{equation}
where $\hat{b}_i^\dagger$ and $\hat{b}_i$ are bosonic creation and annihilation operators for the phonon mode at site $i$, and $\omega_0$ is the phonon frequency.

\paragraph*{Electron--Phonon Interaction.} The Holstein coupling refers a local interaction between electron density and phonon displacement:
\begin{equation}
\hat{H}_{\text{el-ph}} = g \sum_i \left( \hat{b}_i^\dagger + \hat{b}_i \right) \left( \hat{n}_{i\uparrow} + \hat{n}_{i\downarrow} \right),
\label{eq:HH_elph}
\end{equation}
where $g$ denotes the electron-phonon coupling strength. This term generates a phonon-mediated attraction between electrons occupying the same site, counteracting the repulsive \( U \) term.

\subsection{Coherent-State Mean-Field Approximation}
\label{ssec:mf-decoupling}

We describe lattice degrees of freedom employing a coherent-state mean-field approximation for the phonons,\cite{Reinhard2019,Sandhoefer2016} in which the solutions for the bosonic degrees of freedom are assumed to be a coherent state $\ket{\alpha_i}$ for $\alpha_i \in \mathbb{C}$ being a tunable parameter, which satisfies $\hat{b}_i \ket{\alpha_i} = \alpha_i \ket{\alpha_i}$ since coherent state is defined as the eigenstate of the annihilation operator. With this ansatz, the Hamiltonian of the fermi-bose mixtures can be reduced to a fermionic Hamiltonian by substituting the bosonic operators at each site with $\alpha_i$ and $\alpha_i^*$
\begin{equation}
\hat{b}_i \rightarrow \alpha_i, \qquad
\hat{b}_i^\dagger \rightarrow \alpha_i^\ast .
\end{equation}
This approach captures static lattice distortions caused by electron-phonon coupling while maintaining a computationally efficient framework appropriate for large-scale systems.

More specifically, within this approximation, the electron-phonon interaction simplifies to an effective site-dependent potential that influences the electronic subsystem,
\begin{equation}
\hat{H}_{\mathrm{el\text{-}ph}}^{\mathrm{MF}}
= 2g \sum_i \mathrm{Re}(\alpha_i)\,\hat{n}_i ,
\end{equation}
where $\hat{n}_i = \hat{n}_{i\uparrow} + \hat{n}_{i\downarrow}$. The phononic contribution to the energy is correspondingly given by
\begin{equation}
E_{\mathrm{ph}}^{\mathrm{MF}} = \omega_0 \sum_i |\alpha_i|^2 .
\end{equation}

The coherent-state parameter $\{\alpha_i\}$ are obtained through a variational minimization of the total mean-field energy with respect to $\alpha_i^\ast$, resulting in a self-consistency condition that relates the lattice distortion at each site to the local electronic density. This procedure yields a renormalized electronic Hamiltonian that incorporates phonon-induced lattice effects at the mean-field level.

The coherent-state mean-field approximation thus offers a physically justified foundation for the BE approach to treat electronic correlation coupled to phonons, as discussed in the subsequent section. To preserve clarity in the primary text, we only present the fundamental framework of the approximation here. A comprehensive derivation of the mean-field decoupling, the variational equations, and the energy expressions is provided in Appendix~\ref{appendix:meanfield}. In addition, Appendix~\ref{appendix:algorithm} elucidates the variational coherent-state construction and the calculation of phononic contributions to the total ground-state energy.

\subsection{Bootstrap Embedding with Mixed Fermi-Bose Eigensolver for Hubbard-Holstein Model}
\label{ssec:fbbe-alg}

In the BE formalism, the full lattice system is divided into a set of fragments $\{ \mathcal{F}_\alpha \}$, each comprising a small subset of lattice sites. For each fragment, an embedded Hamiltonian $\hat{H}_\alpha^{\mathrm{emb}}$ is constructed by projecting the full Hamiltonian onto the fragment and its associated bath degrees of freedom. The bath orbitals are derived from the mean-field one-particle reduced density matrix (1-RDM) through singular value decomposition, ensuring an accurate representation of the entanglement between the fragment and the environment. \cite{Ye2019JCTC,Liu2023}.

Each embedded Hamiltonian $\hat{H}_\alpha^{\mathrm{emb}}$ is addressed with a precise many-body solver to derive the fragment wavefunctions and associated reduced density matrices, including the fragment 1-RDM $\gamma^{(\alpha)}$, and, when necessary, the two-particle reduced density matrix (2-RDM) $\Gamma^{(\alpha)}$. RDMs across different fragments are not independent; consistency conditions are often enforced within the overlapping regions shared by different fragments to improve the representation of the global many-body state. 
\cite{Liu2023}.

Global consistency is enforced through a Lagrangian formulation,
\begin{equation}
\begin{aligned}
\mathcal{L}
&=
\sum_\alpha E_\alpha
+ \sum_{\alpha < \beta}
  \sum_{ij \in \mathcal{O}_{\alpha\beta}}
  \lambda_{ij}^{(\alpha\beta)}
  \left(
    \gamma_{ij}^{(\alpha)} - \gamma_{ij}^{(\beta)}
  \right) \\
&\quad
- \mu \left(
    \sum_\alpha\sum_i \gamma_{ii} - N_e
  \right),
\end{aligned}
\end{equation}
where $E_\alpha$ represents the ground-state energy of fragment $\alpha$, $\mathcal{O}_{\alpha\beta}$ signifies the overlap sites between fragments $\alpha$ and $\beta$, $\lambda_{ij}^{(\alpha\beta)}$ are Lagrange multipliers enforcing 1-RDM consistency, $\mu$ is a global chemical potential, and $N_e$ is the total number of electrons. Extensions of this formalism to incorporate 2-RDM consistency constraints proceed in an analogous manner\cite{Liu2023}.

The BE equations are solved in a self-consistent manner through iterative updates of the fragment Hamiltonians and Lagrange multipliers until the global RDM constraints reach convergence. In practice, the Lagrange multipliers are regarded as variational parameters and are optimized through the use of gradient-based techniques\cite{Liu2023}.

For the Hubbard–Holstein model examined in this study, the electronic Hamiltonian encompasses the on-site Hubbard interaction and the effective electron–phonon coupling generated by site-local coherent phonon displacements. The phonon degrees of freedom contribute solely a classical energy component, whereas all electronic correlations are addressed expressly within the BE framework. This approach facilitates precise modeling of electron–electron and electron–phonon interactions in large lattice systems that are beyond the capabilities of exact diagonalization.

The BE algorithm to treat Hubbard-Hostein model proceeds as follows, shown in Algorithm \ref{alg:fb-be}. We utilize a CSMF approach for the phonon degrees of freedom integrated with BE for the correlated electronic subsystem. The method iteratively enforces self-consistency between electronic densities, site-local phonon displacements, and global reduced density matrix constraints. 

The Lagrangian formalism of Bootstrap Embedding enforces the global consistency criteria on 1-RDMs and the total particle number. After every cycle of optimizing the fragment 1-RDM, the phonon displacements $\alpha_i$ are updated according to the energy minimization condition:
\[
\alpha_i^\star = -\frac{g}{\omega_0} \langle \hat{n}_i \rangle
\]

 
\begin{figure}[t!]
    \makeatletter 
    \renewcommand{\fnum@figure}{\textbf{Algorithm~\thefigure}} 
    \makeatother

    \hrule height 0.8pt
    \vspace{3pt}
    
    \caption{Self-Consistent Fermi-Bose Bootstrap Embedding (fb-BE)}
    \label{alg:fb-be}
    
    \vspace{3pt}
    \hrule height 0.4pt
    
    \begin{algorithmic}[1]
        \State \textbf{Input:} Number of Sites $N$, electrons number $N_e$, Hubbard interaction $U$, phonon frequency $\omega_0$, electron-phonon coupling $g$, tolerances $\varepsilon_{\text{inner}}, \varepsilon_{\text{outer}}$
        \State Initialize $\{\alpha_i^{(0,0)}\} \in \mathbb{C}$
        
        \For{grand iteration $m = 0$ to max\_iter}
            \State Set $\alpha_i^{(m,0)} = \alpha_i^{(m)}$
            
            \For{inner iteration $k = 0$ to max\_inner}
                \State Construct electronic Hamiltonian:
                \State $\displaystyle h_{ij} = 
                \begin{cases}
                    -t & \text{if } i,j \text{ are nearest neighbors} \\
                    2g \cdot \text{Re}(\alpha_i^{(m,k)}) & \text{if } i=j
                \end{cases}$
                
                \State Solve RHF with on-site interaction $U$ and effective coupling term
                \State Extract site densities $n_i^{(m,k)}$ from 1-RDM
                \State Update $\alpha_i^{(m,k+1)} = -\frac{g}{\omega_0} n_i^{(m,k)}$
                
                \If{$\|\alpha^{(m,k+1)} - \alpha^{(m,k)}\| < \varepsilon_{\text{inner}}$}
                    \State \textbf{Break} inner loop
                \EndIf
            \EndFor
            
            \State 
            Construct fragment and bath orbitals from the converged mean-field solution via Schmidt decomposition
            also construct embedded fragment hamiltonian
            \State 
            Solve fragment hamiltonian using FCI and perform BE optimization
            \State Extract BE 1-RDM $\gamma_{ij}^{m}$ and compute $n_i^{\text{BE},(m)} = \gamma_{ii}^{m}$
            \State Update: $\displaystyle \alpha_i^{(m+1)} = -\frac{g}{\omega_0} n_i^{\text{BE},(m)}$
            
            \If{$\|\alpha^{(m+1)} - \alpha^{(m)}\| < \varepsilon_{\text{outer}}$}
                \State \textbf{Break} grand loop
            \EndIf
        \EndFor
        \State \textbf{Output:} Final $\{\alpha_i\}$, $E_{\text{tot}} = E_{\text{el}} + \omega_0 \sum_i |\alpha_i|^2$
    \end{algorithmic}
    
    \vspace{2pt}
    \hrule height 0.8pt
\end{figure}


\section{Computational Details}
\label{sec:comp}
We study the one-dimensional Hubbard–Holstein model with nearest-neighbor hopping $t$. All energies are therefore expressed in units of $t=1$. The phonon frequency is fixed at $\omega_0 = 0.25$. 

For benchmarking, we first consider an eight-site chain at both half filling ($N_e = 8$) and quarter filling ($N_e = 4$). The electron–phonon coupling strength $g$ is varied from $0.1$ to $0.5$, and the on-site Coulomb repulsion $U$ is varied from $0$ to $10$. To assess scalability, we  consider larger half-filled chains containing up to 350 lattice sites at $g=0.1$.
The DMRG calculation is performed with bond-dimension=150 using the Block2 package \cite{block2} via computational resources provided by the ACCESS program \cite{access}. In the DMRG calculation, each phonon site was truncated to 11 Fock levels with local bosonic Hilbert space of dimension 11, corresponding to a maximum phonon occupation number 10 per lattice site. fb-BE calculations were performed using the customized version of  QuEmb package \cite{quemb,github_quemb}.

We use the BE2 \cite{quemb} scheme to break up the full lattice system into a number of overlapping fragments. In this scheme, we fragment the system by looping over all spatial lattice sites, generating fragments that consist of that central site and its nearest neighbors (i.e., a fragment radius of 1 site). We leave out the first and last fragments to avoid edge overcounting. To build the embedded Hamiltonian $\hat{H}_\alpha^{\text{emb}}$  for each fragment $\mathcal{F}_\alpha$, we use the mean-field-decoupled Hubbard–Holstein form for the global system. The detailed derivation can be found in Appendix~\ref{appendix:meanfield}. 

Exact diagonalization is used to solve each embedded Hamiltonian. This gives us fragment reduced density matrices (RDMs). The BE optimization's convergence threshold (the consistency of the fragment and global reduced density matrices) was set on density mismatch at $10^{-6}$. To provide self-consistent coupling between phonon displacements and electronic structure, two nested loops were utilized: an inner Hubbard–Holstein mean-field loop and a grand outer BE-coupled loop. The convergence of the inner electronic–phonon loop occurred when the $\ell_2$-norm variation in the phonon displacement fields $\{\alpha_i\}$ decreased below $10^{-5}$. The outer self-consistency loop, which linked BE-derived site occupations to new phonon displacements, converged when the $\ell_2$-norm difference in $\{\alpha_i\}$ between two iterations was less than $5 \times 10^{-6}$. These requirements made sure that both the electronic and bosonic degrees of freedom converged completely in all calculations.

\section{Results and Discussion}
\label{sec:results}

In this section, we present numerical results obtained using the fb-BE framework for the one-dimensional Hubbard–Holstein model. In Sec.~\ref{result:conv}, we show the convergence of $\alpha$ values for the outer iteration of the algorithm and the convergence of the fragment-size. In Sec.~\ref{resul:scaling}, finite-size scaling to infinite-size limit is presented, and a runtime comparison to DMRG is performed. In Sec.~\ref{result:e-p physics}, we discuss the electron-phonon physics for both half-filled and quarter-filled cases for 8-site system and compared our result with DMRG.

\subsection{Convergence of the Algorithm}
\label{result:conv}
The outer Hubbard--Holstein self-consistency is implemented as a fixed-point iteration of updating $\alpha$. We also used linear mixing strategy to update $\alpha$ where it faced challenge to get converged, especially in the strong coupling region for some values of U.
Convergence is monitored using the Euclidean norm of the update
\begin{equation}
\left\| \Delta \alpha^{(m)} \right\|_2 
= \left\| \alpha^{(m+1)} - \alpha^{(m)} \right\|_2,
\end{equation}
with a tolerance of $10^{-6}$. In all parameter regimes considered, the residual decreases monotonically and approximately exponentially with iteration number. Representative convergence behavior is shown in Fig.~\ref{fig:alpha_convergence}.
\setcounter{figure}{0}
\begin{figure}[htbp!]
\centering
\includegraphics[width=\linewidth]{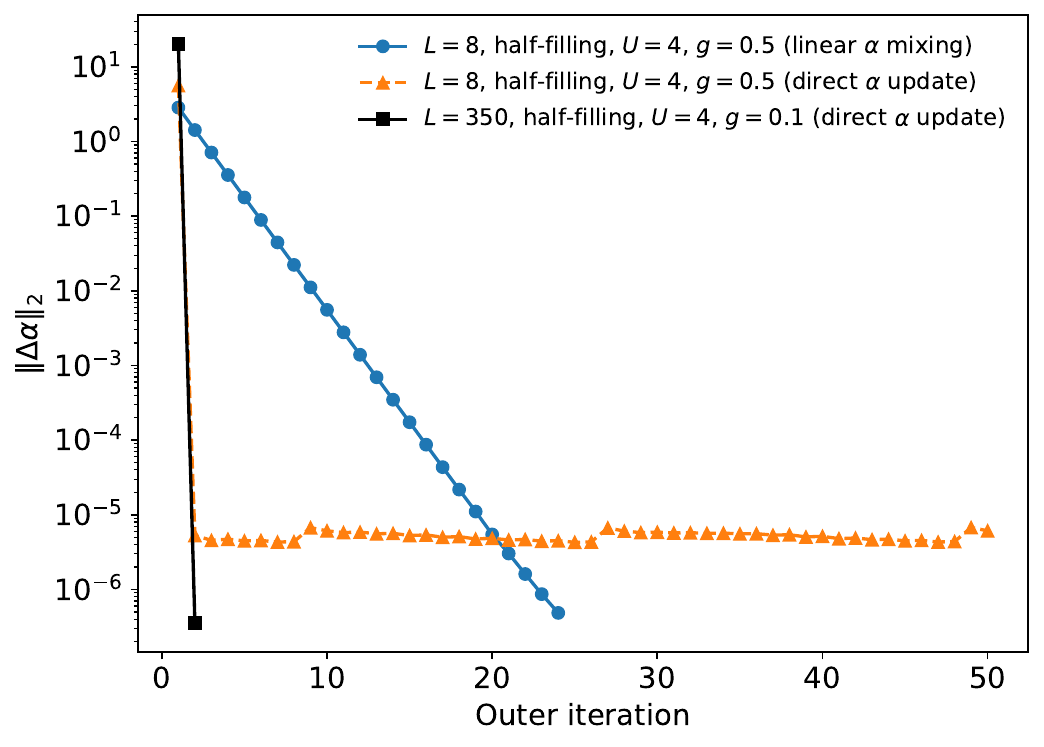}
\caption{Convergence of the outer Hubbard--Holstein self-consistency loop, quantified by the Euclidean norm of the phonon displacement update 
$\|\Delta \alpha^{(k)}\|_2$ as a function of outer iteration. 
For the $L=8$ half-filled system at $U=5$, $g=0.5$ (blue circles), linear mixing of $\alpha$ produces monotonic and approximately exponential convergence. 
For the larger system ($L=350$, $U=5$, $g=0.1$; orange squares), the direct (unmixed) update converges in two iterations, with the residual dropping below $10^{-6}$ after the first correction.}
\label{fig:alpha_convergence}
\end{figure}

For the $L=8$ half-filled system at $U=5$ and $g=0.5$, the residual decreases monotonically and approximately exponentially with iteration number, reaching the convergence threshold within 25 iterations because of the linear mixing scheme but it got stuck by direct update of $\alpha$ values. 
For the larger system ($L=350$, $U=5$, $g=0.1$), the direct (unmixed) update converges in two iterations, with the residual reduced below $10^{-6}$ after the first correction. 
To assess convergence with respect to fragment size, we compared BE2, and BE3 for an 8-site chain at half filling in the strong-coupling regime ($U = 8$, $g = 0.5$, $t = 1$). The total energies are
\[
 \quad
E_{\mathrm{tot}}^{\mathrm{BE2}} = -10.4200, \quad
E_{\mathrm{tot}}^{\mathrm{BE3}} = -10.4208.
\]
The BE2 and BE3 results differ by $8 \times 10^{-4}$, indicating that BE2 is essentially converged for this small system.

\subsection{Finite-Size Scaling to Infinite Size}
\label{resul:scaling}
To assess finite-size effects and establish convergence toward bulk behavior of our fb-BE method, we performed a systematic finite-size scaling analysis of the ground-state energy density, energy per unit lattice site, for half-filled 1D systems with lattice sizes $L = 150, 200, 300,$ and $350$ at coupling strength $g=0.1$ and phonon frequency $\omega_0=0.25$.

\begin{figure}[htbp!]
\centering
\includegraphics[width=\linewidth]{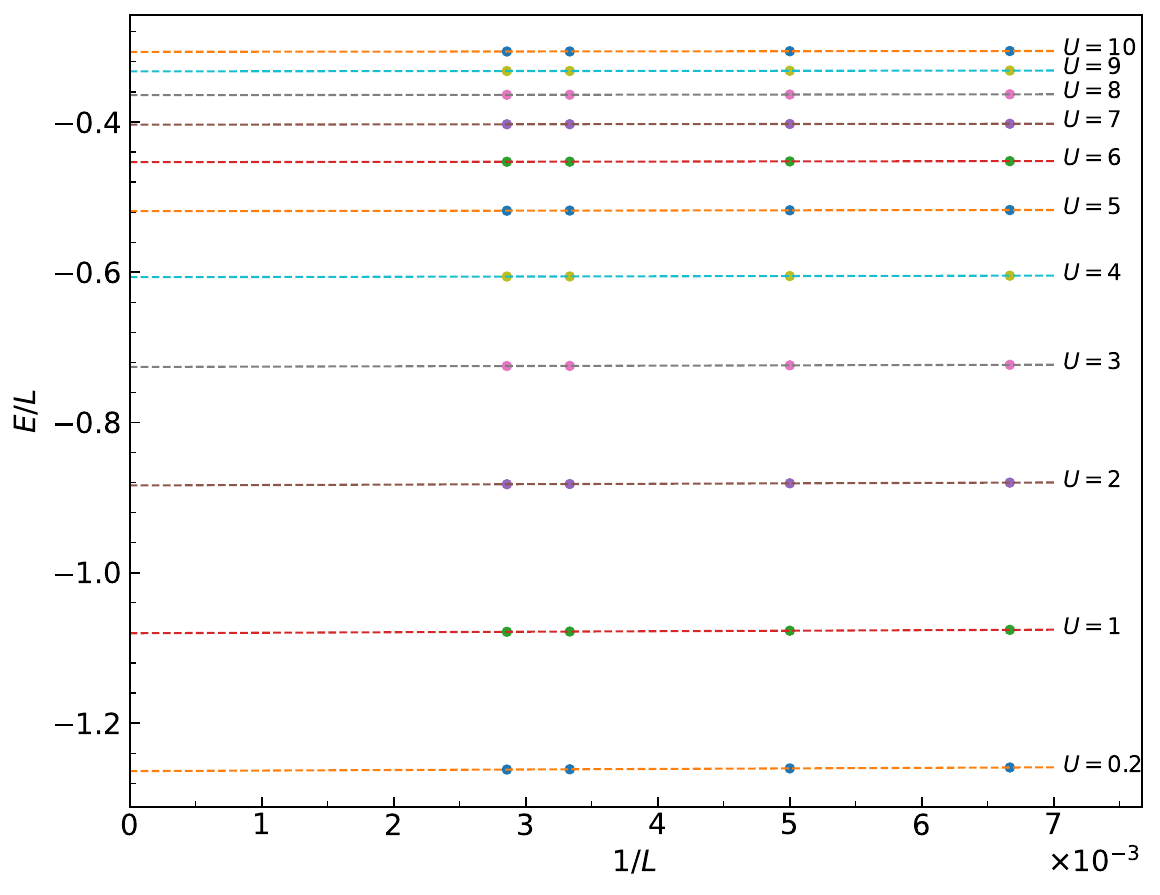}
\caption{
Finite-size scaling of the ground-state energy per site at at different $U=$ values. The energy density $E(L,U)$ is plotted versus inverse system size $1/L$. The dashed line is a linear fit of the form $E(L,U)=E_\infty(U)+a(U)/L$, whose intercept yields the infinite-size-limit value $E_\infty$.
}
\label{fig:scaling_example}
\end{figure}

Fig. ~\ref{fig:scaling_example} shows the scaling behavior at different interaction strength $U=4$. The energy per site exhibits an essentially linear dependence on $1/L$, consistent with leading finite-size corrections of the form
\begin{equation}
E(L,U) = E_\infty(U) + \frac{a(U)}{L}.
\end{equation}
 The intercept of the linear fit at $1/L \to 0$ provides the infinite-size energy density $E_\infty(U)$. The  linearity confirms that finite-size effects are well described by leading $1/L$ scaling over the system sizes considered. We perform such finite-size scaling for different $U$, and the final per-site energy extrapolated to infinite size limit is tabulated in Table \ref{tab:einf_scaling}.


\begin{table}[htbp]
\caption{Infinite-size-limit ground-state energy per site $E_\infty(U)$
obtained from linear extrapolation in $1/L$ using system sizes
$L = 150, 200, 300,$ and $350$ at $g = 0.1$ and $\omega_0 = 0.25$.}
\label{tab:einf_scaling}
\centering
\begin{tabular}{c|cccccc}
\hline 
$U$ & 0.2 & 1.0  &2.0  &3.0  &4.0  & 5.0  \\
\hline
 $E_\infty(U)$ & $-1.264$ & $-1.080$ & $-0.884$ & $-0.726$ & $-0.607$ & $-0.519$ \\
\hline
\multicolumn{7}{c}{\vspace{0.05cm}}\\
\end{tabular}
\begin{tabular}{c|ccccc}
\hline
$U$ & 6.0  &7.0  & 8.0  & 9.0  & 10.0 \\
\hline
 $E_\infty(U)$ & $-0.454$ & $-0.404$ & $-0.364$ & $-0.333$  & $-0.307$ \\
\hline
\end{tabular}
\end{table}

\begin{figure}[t]
\centering
\includegraphics[width=\linewidth]{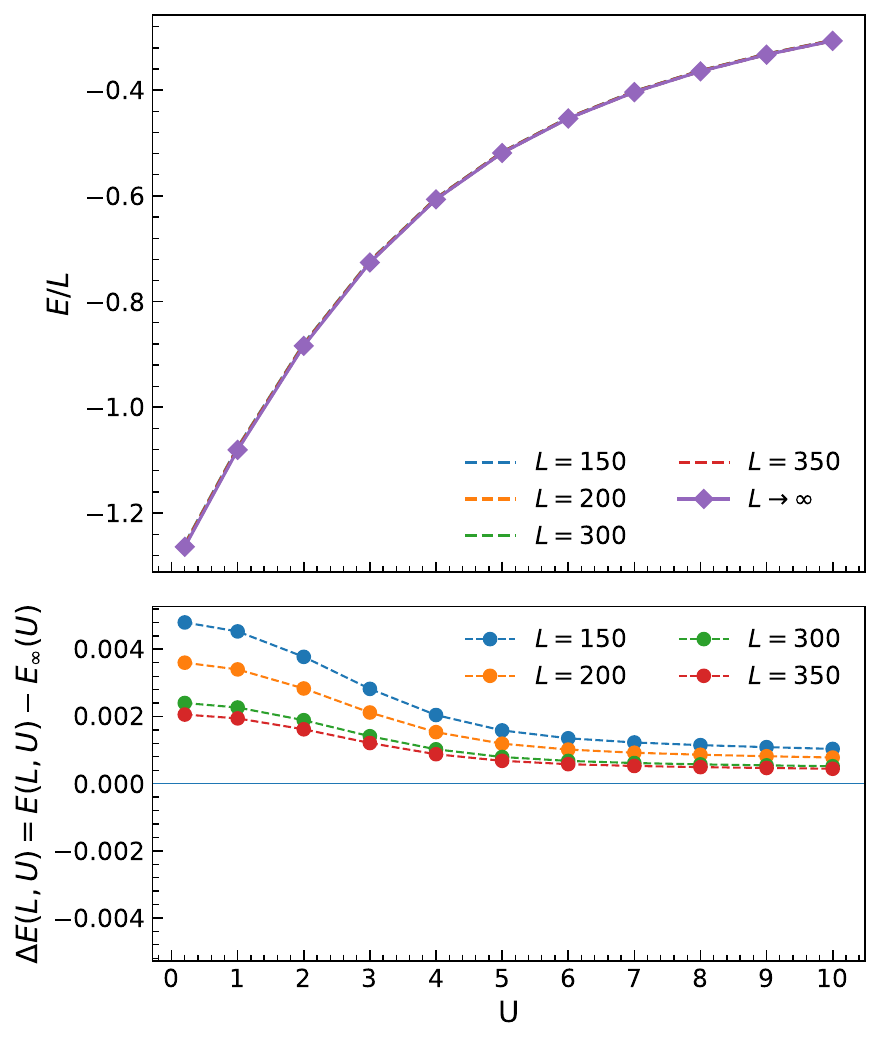}
\caption{
Ground-state energy per site $E(L,U)$ as a function of interaction strength $U$ for $L=150$--350 together with the extrapolated infinite-size result $E_\infty(U)$ (upper panel). 
Finite-size deviation $\Delta E(L,U)=E(L,U)-E_\infty(U)$, highlighting the systematic suppression of size effects with increasing $L$ (lower panel).
}
\label{fig:scaling_full}
\end{figure}

To see explicitly the dependence of the energy per-site on interaction strength, Fig.~\ref{fig:scaling_full} shows the deviation of per-site energy for finite-size system as compared to the extrapolated infinite-size limit. The upper panel shows $E(L,U)$ for all system sizes together with the extrapolated infinite-size curve $E_\infty(U)$. The finite-size results converge monotonically toward the bulk limit across the entire interaction range. In particular, the $L=300$ and $L=350$ curves are nearly indistinguishable from $E_\infty(U)$, indicating that lattice sizes $L \gtrsim 300$ already provide an accurate approximation to infinite-size behavior.

To quantify the residual size dependence, the lower panel of Fig.~\ref{fig:scaling_full} displays the finite-size deviation
\begin{equation}
\Delta E(L,U) = E(L,U) - E_\infty(U).
\end{equation}
The deviations decrease systematically with increasing system size and remain small over the full interaction range. The largest deviations occur at smaller $U$ while the deviations diminish at larger $U$, consistent with the increasingly local character of the strongly interacting regime. Overall, the nearly linear $1/L$ scaling and the rapid suppression of $\Delta E(L,U)$ demonstrate that the fb-BE calculations are well converged with respect to system size and reliably capture the infinite-size energy density for $L \ge 300$.

\begin{figure}[htbp!]
\centering
\includegraphics[width=\linewidth]{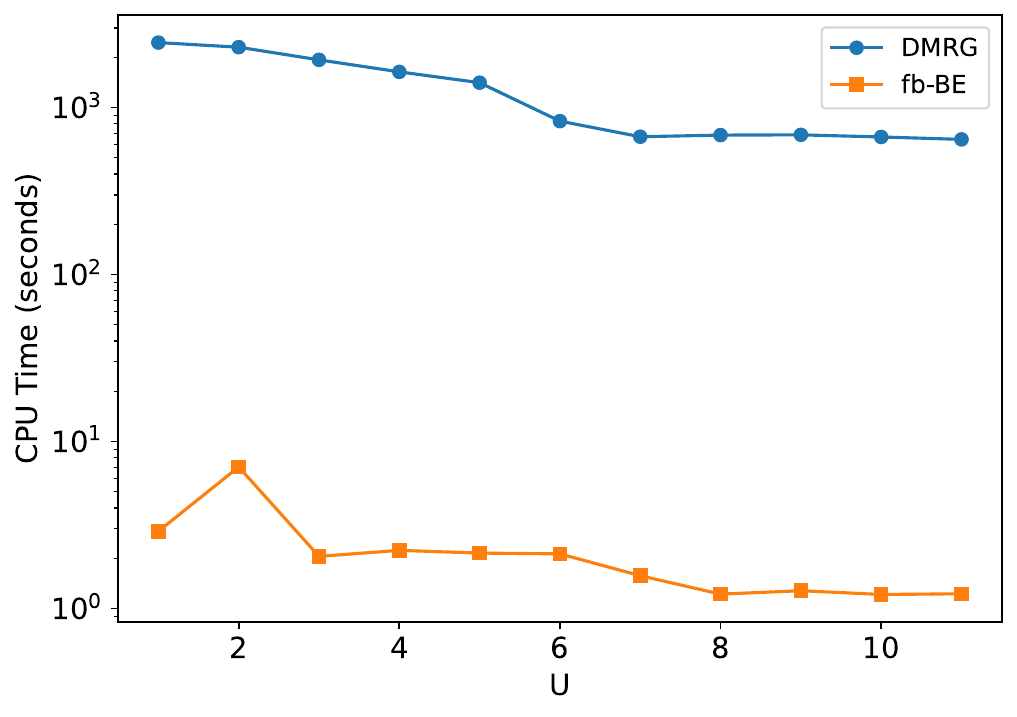}
\caption{CPU time (total runtime) comparison between DMRG and fb-BE for an 8-site Hubbard--Holstein chain at half filling ($g=0.4$).}

\label{fig:cpu_time_comparison}
\end{figure}
Fig.~\ref{fig:cpu_time_comparison} shows the CPU time required by DMRG and fb-BE across the interaction sweep. 
The embedding approach is consistently faster. 
While DMRG requires $\mathcal{O}(10^{3})$ seconds per $U$ value, fb-BE converges within a few seconds and displays only weak dependence on interaction strength. 
These results highlight the substantial computational efficiency of the embedding framework.

\subsection{Probing Electron-Phonon Physics}
\label{result:e-p physics}

In this part, we compare the performance of the fb-BE method's with DMRG calculations at half-filling in Sec. \ref{ssec:hf} and quater-filling in Sec.~\ref{ssec:qf} for a small lattice of 8 sites. The small system size is chosen mostly due to the DMRG calculation efficiency. We note similar calculation can be performed for larger size systems given enough computational time.

\subsubsection{Half-Filling}
\label{ssec:hf}
We start by looking at the half-filled case, where the physics is determined by the interaction between the charge-density-wave (CDW) instability caused by $g$ and the Mott insulating state affected by $U$ \cite{Costa2020, Clay2005}.

\begin{figure}[h!]
    \centering
    \includegraphics[width=\linewidth]{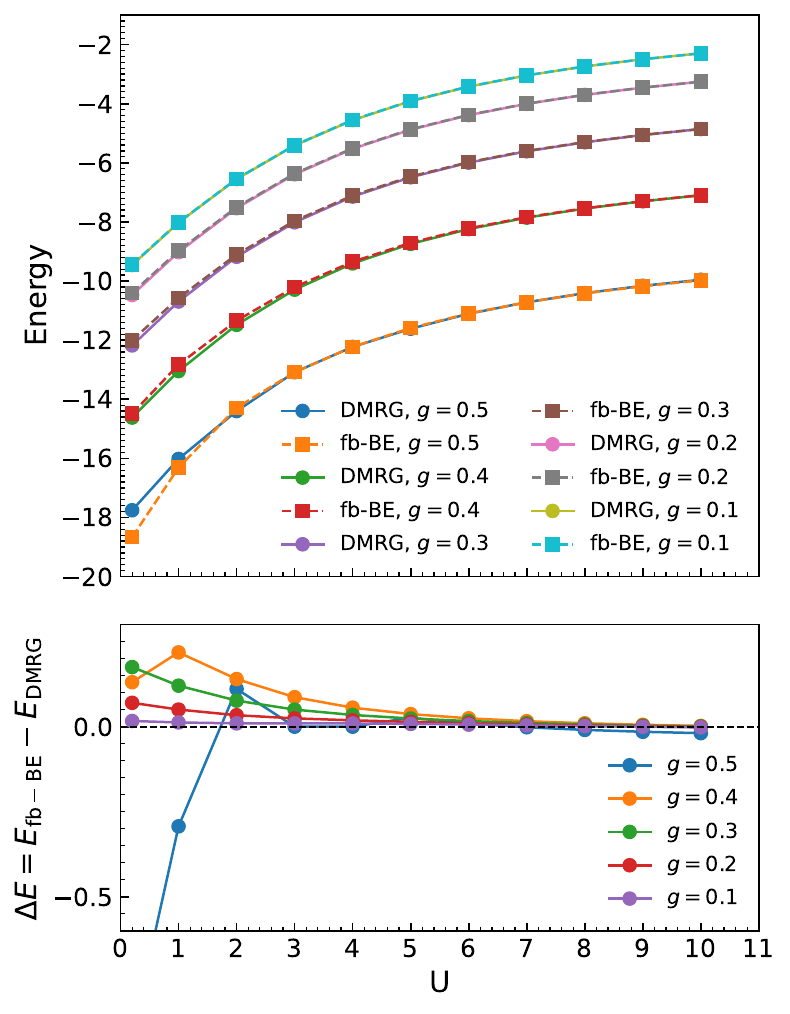}
    \caption{Ground state energy comparison between DMRG and fb-BE for an 8-site, 8-electron system ($\omega_0=0.25$). The embedding method accurately captures the energy evolution from the weak-coupling to strong-coupling regimes. The substantial negative deviation observed for low $U$ and high $g$ is attributed to the mean-field overstabilization of the static CDW order; however, this deviation disappears in the Mott regime.}
    \label{fig:8e_energy}
\end{figure}

Fig.~\ref{fig:8e_energy} shows the total ground state energies obtained from DMRG and fb-BE. The method shows great qualitative agreement all around the phase diagram. As expected, the overall energy goes up (becomes less negative) when the repulsive $U$ goes up. On the other hand, stronger electron-phonon coupling lowers the energy significantly. 
The reduction in energy results from lattice relaxation: the lattice deforms to shield the local electronic density, thereby screening the Coulomb interaction and producing a phonon-mediated attractive interaction that counteracts the repulsive Hubbard term\cite{Holstein1959}.

The lower panel of Fig.~\ref{fig:8e_energy} depicts the energy deviation $\Delta E = E_{\text{fb-BE}} - E_{\text{DMRG}}$ to further quantify the correctness of the approach. There are two different regimes to be seen. \textbf{Strong Correlation Regime ($U > 2t$):} The errors are essentially independent of the coupling strength $g$ and are very modest ($|\Delta E| < 0.05$ Hartree). Electrons are confined and charge fluctuations are controlled in this Mott-dominated regime \cite{Costa2020}. Here, the short-range antiferromagnetic interactions are well captured by the local fragment logic of the Bootstrap Embedding framework \cite{Ye2019JCTC}. \textbf{Peierls/CDW Regime for large $g$, when $U < 2t$:} There are notable variations, especially for $g=0.5$ at low $U$, where $\Delta E \approx -0.5$ Hartree. The fb-BE energy is less than the DMRG energy, as shown by the negative sign. The CSMF method doesn't take into account quantum phonon fluctuations and instead assumes a static lattice distortion, which is a known problem with the mean-field treatment of phonons and the negative energy error here can likely be attributed to the non-variational nature of bootstrap embedding. Quantum fluctuations (zero-point motion and tunneling between degenerate CDW patterns) increase the ground state energy in the exact solution (DMRG)\cite{Jeckelmann1999,QWang1999}. Because the mean-field solution essentially "freezes" the system into a deeper potential well, the binding energy is overestimated.


\subsubsection{Quarter-Filling}
\label{ssec:qf}
\begin{figure}[h!]
    \centering
    \includegraphics[width=\linewidth]{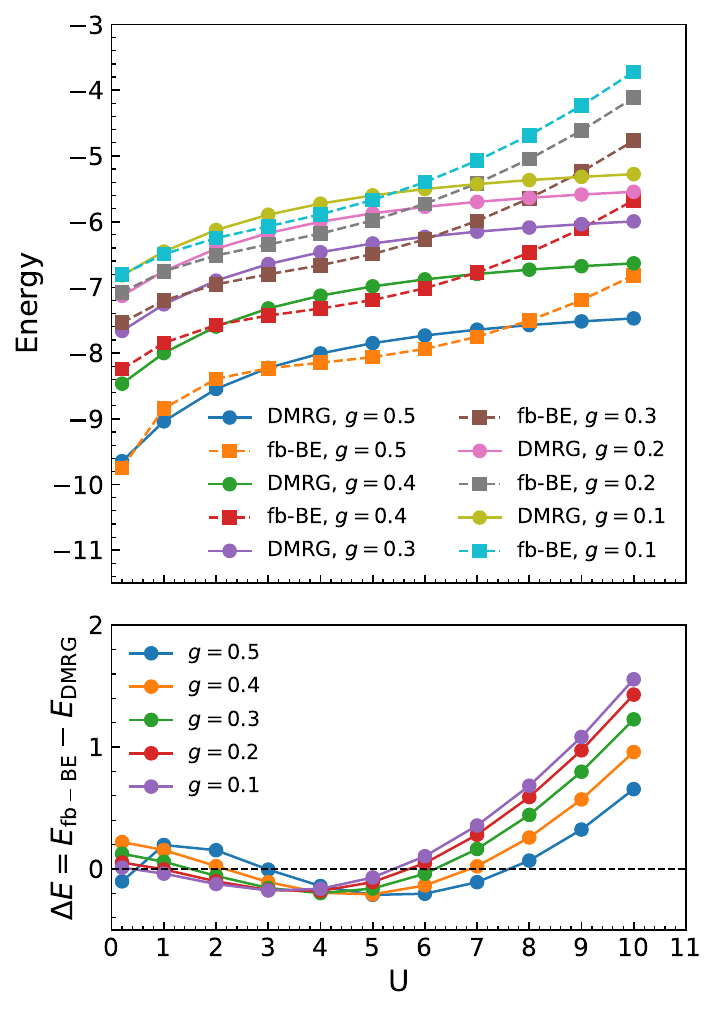}
    \caption{The energy of the ground state for the quarter-filled system (8 sites, 4 electrons). When $U$ is big, the differences between fb-BE and DMRG are more than when it is half-filled.Energy deviation for the system that is a quarter full ($N_e=4$). The inaccuracy is more as $U$ gets bigger, but it gets smaller when $g$ gets stronger. This means that phonon-induced localization (polaron production) makes the local embedding approximation more accurate.}
    \label{fig:4e_energy}
\end{figure}

Next, we look at the system with $N_e = 4$ (quarter-filling). This regime usually has either metallic behavior or a liquid of polarons, depending on how strong the coupling is \cite{Hardikar2007}. The results are shown in Fig.~\ref{fig:4e_energy}.


In this situation, the errors are mostly positive ($E_{\text{fb-BE}} > E_{\text{DMRG}}$), which means that the embedding ansatz is missing some of the correlation energy. The bottom panel of Fig.~\ref{fig:4e_energy} shows a surprising change in accuracy caused by the interaction between electrons and phonons. \textbf{Weak Coupling ($g=0.1$):} The error increases monotonically with $U$ for $U \gtrsim 4$, reaching $\Delta E \approx 1.5$ Hartree at $U=10$. In this quarter-filled regime, the system retains its metallic characteristics despite significant repulsion, as the holes facilitate charge mobility without double occupancy. We classify this as a strongly correlated metallic phase \cite{Hardikar2007}. The small fragment size utilized in BE captures on-site correlations well but difficulties with the longer-range kinetic correlations of delocalized electrons.

\textbf{Strong Coupling ($g=0.5$):} The error is significantly lower than for weak coupling ($\Delta E \approx 0.6$ Hartree at $U=10$). Within the CSMF approximation, phonon degrees of freedom are treated as static classical displacements. In the strong-coupling regime, the electron–lattice interaction generates spatially inhomogeneous, self-consistent lattice distortions that renormalize the local on-site energies. These site-dependent effective potentials enhance charge localization and reduce the spatial extent of electronic correlations.

As a result, the effective electronic Hamiltonian becomes more local, which is favorable for bootstrap embedding since the fragmentation error is minimized when correlations are predominantly short-ranged. The improved performance of fb-BE at strong coupling is therefore attributed to enhanced electronic localization induced by static lattice distortions. In particular, at large $g$, the lattice-induced potential can promote partial spatial charge separation, which reduces the degree of electronic delocalization. The continued variation of the total energy with increasing
$U$ should therefore be interpreted as residual electronic correlation effects within this localized regime.




\section{Conclusion}
\label{sec:conclu}
We propose a new fermi-bose Bootstrap Embedding (fb-BE) method and demonstrated its convergence using one-dimensional Hubbard–Holstein model for large system size. In addition, systematically benchmarks against precise DMRG calculations are performed for small system size. We have explored the applicability domain of this embedding strategy by investigating two unique filling regimes (half and quater filling) that exhibit different physics.

Our results show that fb-BE works very well when there is a large correlation ($U > 2t$), because electron localization reduces the errors that come with finite fragment sizes and it is consistent with the findings from DMET applied to Hubbard-Holstein model \cite{Reinhard2019}. At half-filling, the technique precisely represents the Mott insulating state, exhibiting minimal divergence from DMRG over a broad spectrum of electron-phonon coupling strengths. In the weak-correlation Peierls regime ($U < 2t$, large $g$), the static mean-field approach to phonons results in an exaggerated assessment of charge-density-wave stability, underscoring the importance of incorporating quantum phonon fluctuations to accurately determine the ground state energy near the phase transition.

In the quarter-filling regime, we noted a distinct transition in accuracy induced by polaron production. The approach fails to accurately capture the long-range kinetic correlations of delocalized electrons (weak $g$), but the emergence of strong electron-phonon coupling markedly enhances the outcomes. The creation of small polarons confines the electronic degrees of freedom, thus improving the accuracy of the local embedding approximation.

The fb-BE method offers a computationally efficient method for modeling strongly correlated electron-phonon systems, especially in regimes characterized by localizing interactions. Future developments of this research may mitigate existing constraints by integrating dynamical phonon treatments or enlarging the fragment size to more effectively capture long-range changes in the metallic phase.


Our work also opens several directions for future exploration. One is to improve the quality of the mixed fermi-bose eigensolver by including phonon correlation beyond a single coherent state \cite{Li2024PRL}. However, classical eigensolvers will face a bottleneck for strongly correlated systems at large size. One hopeful approach is to use quantum computers for fragment eigensolvers \cite{Liu2023,Bierman2025ChemRxiv,Bidart2025arxiv,Joven2025arxiv}. In addition, because the electron-phonon problem has both discrete fermionic and continuous bosonic degrees of freedom, hybrid continuous-variable (CV) and discrete-variable (DV) quantum processors \cite{Andersen2015,Joven2025arxiv,Kemper2025arxiv2511,Liu2024arxiv} could be an excellent choice to use as a quantum eigensolver. Ultimately, extending the current formalism to include excited states and finite-temperature phase diagrams would significantly enhance our understanding of the thermodynamics of polaronic materials \cite{Jankovic2022}.

\section*{Acknowledgments}
This work is supported by the U.S. Department of Energy, Office of Science, Advanced Scientific Computing Research, under contract number DE-SC0025384. This work used Bridges-2 at the Pittsburgh Supercomputing Center (PSC) through allocation CHE250139 from the Advanced Cyberinfrastructure Coordination Ecosystem: Services \& Support (ACCESS) program, which is supported by U.S. National Science Foundation grants \#2138259, \#2138286, \#2138307, \#2137603, and \#2138296.

\section*{Data Availability}
The data that support the findings of this study are available in the fb-BE github repository\cite{fbBE_2025}.

\appendix
\section{Effective Hamiltonian for the Hubbard-Holstein Model}
\label{appendix:meanfield}

We consider the Hubbard--Holstein model on a lattice including $N$ sites, wherein electrons undergo  an on-site Coulomb interaction and local coupling to quantized phonon modes. The full Hamiltonian is:
\begin{equation}
\begin{aligned}
\hat{H} = -t \sum_{\langle i, j \rangle, \sigma} \left( \hat{c}_{i\sigma}^\dagger \hat{c}_{j\sigma} + \text{h.c.} \right)
+ U \sum_i \hat{n}_{i\uparrow} \hat{n}_{i\downarrow}
+ \omega_0 \sum_i \hat{b}_i^\dagger \hat{b}_i
\\+ g \sum_{i,\sigma} \hat{n}_{i\sigma} \left( \hat{b}_i^\dagger + \hat{b}_i \right),
\end{aligned}
\end{equation}
where $\hat{c}_{i\sigma}^\dagger$ ($\hat{c}_{i\sigma}$) denotes the creation (annihilation) an electron with spin $\sigma$ at site $i$, and $\hat{n}_{i\sigma} = \hat{c}_{i\sigma}^\dagger \hat{c}_{i\sigma}$ is the corresponding number operator. The phonon operators $\hat{b}_i^\dagger$, $\hat{b}_i$ create and annihilate vibrational quanta at site $i$. The parameters $t$, $U$, $\omega_0$, and $g$ denote the electron hopping amplitude, on-site Coulomb repulsion, phonon frequency, and electron--phonon coupling strength, respectively.

\subsection*{Mean-Field Decoupling}

To achieve a manageable approximation, we implement mean-field decoupling on the interaction terms. The Hubbard term is expressed in a linearized form as follows:
\begin{equation}
    \hat{n}_{i\sigma} =
    \langle\hat{n}_{i\sigma}\rangle + \left( \hat{n}_{i\sigma}- \langle \hat{n}_{i\sigma}\rangle\right) 
    =\langle\hat{n}_{i\sigma}\rangle + \delta\hat{n}_{i\sigma}
\end{equation}
so,
\begin{equation}
    \hat{n}_{i\uparrow}\hat{n}_{i\downarrow}=\left(\langle\hat{n}_{i\uparrow}\rangle + \delta\hat{n}_{i\uparrow}\right)
\left(\langle\hat{n}_{i\downarrow}\rangle + \delta\hat{n}_{i\downarrow}\right)
\end{equation}
Neglecting the second order term $\delta\hat{n}_{i\uparrow}\delta\hat{n}_{i\downarrow}$ because it is a product of two small fluctuations,

\begin{equation}
\hat{n}_{i\uparrow} \hat{n}_{i\downarrow} \approx \langle \hat{n}_{i\uparrow} \rangle \hat{n}_{i\downarrow}
+ \langle \hat{n}_{i\downarrow} \rangle \hat{n}_{i\uparrow}
- \langle \hat{n}_{i\uparrow} \rangle \langle \hat{n}_{i\downarrow} \rangle.
\end{equation}

Similarly, the electron--phonon term is decoupled as follows:

\begin{equation}
\hat{n}_{i\sigma} \left( \hat{b}_i^\dagger + \hat{b}_i \right) \approx
\langle \hat{n}_{i\sigma} \rangle \left( \hat{b}_i^\dagger + \hat{b}_i \right)
+ \hat{n}_{i\sigma} \langle \hat{b}_i^\dagger + \hat{b}_i \rangle
- \langle \hat{n}_{i\sigma} \rangle \langle \hat{b}_i^\dagger + \hat{b}_i \rangle.
\end{equation}

The full mean-field Hamiltonian becomes:
\begin{equation}
\begin{aligned}
\hat{H}^{\text{MF}}_{\text{el-ph}} = \ 
& -t \sum_{\langle i,j \rangle, \sigma} \left(\hat{c}^\dagger_{i,\sigma} \hat{c}_{j,\sigma} +\text{h.c.}\right) 
\\&+ U \sum_i \left( \langle \hat{n}_{i,\downarrow} \rangle \hat{n}_{i,\uparrow} 
+ \langle \hat{n}_{i,\uparrow} \rangle \hat{n}_{i,\downarrow} \right) \nonumber \\
& + \sum_i \omega_0 \hat{b}^\dagger_i \hat{b}_i 
+ g \sum_{i,\sigma} \left( \langle \hat{n}_{i\sigma} \rangle (\hat{b}^\dagger_i + \hat{b}_i) 
+ \hat{n}_{i\sigma} \langle \hat{b}^\dagger_i + \hat{b}_i \rangle \right) \nonumber \\
& - U \sum_i \langle \hat{n}_{i,\uparrow} \rangle \langle \hat{n}_{i,\downarrow} \rangle 
- g \sum_{i,\sigma} \langle \hat{n}_{i\sigma} \rangle \langle \hat{b}^\dagger_i + \hat{b}_i \rangle
\end{aligned}
\end{equation}

\subsection*{Total Effective Hamiltonian}
The resulting total Hamiltonian is
\begin{equation}
    \hat{H}_{\text{el-ph}}^{\text{MF}} = \hat{H}_{\text{el}}^{\text{MF}} \otimes \mathbb{I}_{\text{ph}} + \mathbb{I}_{\text{el}} \otimes \hat{H}_{\text{ph}}^{\text{MF}},
\end{equation}
where the fermionic part is
\begin{align}
    \hat{H}_{\text{el}}^{\text{MF}} = -t \sum_{\langle i, j \rangle, \sigma} \left( \hat{c}_{i\sigma}^\dagger \hat{c}_{j\sigma} + \text{h.c.} \right) \nonumber \\
    + \sum_{i,\sigma} \left( U \langle \hat{n}_{i\bar{\sigma}} \rangle + g \langle \hat{b}_i + \hat{b}_i^\dagger \rangle \right) \hat{n}_{i\sigma}
\end{align}
and the bosonic part is
\begin{equation}
    \hat{H}_{\text{ph}}^{\text{MF}} = \sum_i \left[
\omega_0 \hat{b}_i^\dagger \hat{b}_i + g \langle \hat{n}_{i\sigma} \rangle \left( \hat{b}_i^\dagger + \hat{b}_i \right)
\right].
\end{equation}
Despite the name, we emphasize that $\hat{H}_{\text{el}}^{\text{MF}}$ is still an interacting correlated Hamiltonian, which will be solved using correlated eigensolvers. The superscript "MF" only refers to the interaction between fermions and bosons are treated at the mean-field level.

Assuming the phonon state is a product of coherent states $\ket{\alpha_i}$ which satisfy the following relationships
\begin{align*}
\hat{b}_i \ket{\alpha_i} = \alpha_i \ket{\alpha_i}, ~~
\langle \hat{b}_i^\dagger \hat{b}_i \rangle = |\alpha_i|^2, ~~
\langle \hat{b}_i + \hat{b}_i^\dagger \rangle = 2\,\text{Re}(\alpha_i).
\end{align*}

Define shifted displaced operators
\begin{align}
    \hat{b}_i' = \hat{b}_i + \frac{g \langle \hat{n}_{i\sigma} \rangle}{\omega_0}, \quad
\hat{b}_i'^\dagger = \hat{b}_i^\dagger + \frac{g \langle \hat{n}_{i\sigma} \rangle}{\omega_0}, 
\end{align}
the phonon Hamiltonian can be rewritten in terms of the new shifted bosonic operators as
\begin{align}
    \hat{H}_{\text{ph}}^{\text{MF}} = \sum_i \left[ \omega_0 \hat{b}_i'^\dagger \hat{b}_i' - \frac{g^2 \langle \hat{n}_{i\sigma} \rangle^2}{\omega_0} \right].
\end{align}
As a special case, if the bosonic ground state is vacuum, i.e., $\langle\hat{b}_i'^\dagger\hat{b}_i'\rangle = 0$, then the phonon energy is
\begin{equation}
E_{\text{ph}}=-\frac{\sum_i g^2\langle\hat{n}_{i\sigma}\rangle^2}{\omega_0} .
\end{equation}

\subsection*{Total Effective Energy}
\begin{equation}
E_{\text{MF}} = E_{\text{el}} + E_{\text{ph}} 
- g \sum_{i,\sigma} \langle \hat{n}_{i\sigma} \rangle \langle \hat{b}_i + \hat{b}_i^\dagger \rangle
- U \sum_i \langle \hat{n}_{i\uparrow} \rangle \langle \hat{n}_{i\downarrow} \rangle.
\end{equation}

\section{Variational Coherent-State Ansatz as Mixed Fermi-Bose Eigensolver for the Hubbard-Holstein Model}
\label{appendix:algorithm}
We consider the Hubbard--Holstein Hamiltonian:
\begin{equation}
\hat{H} = \hat{H}_{\text{el}} + \sum_i \omega_0 \hat{b}_i^\dagger \hat{b_i} + g \sum_i \hat{n_i} (\hat{b}_i^\dagger + \hat{b_i}),
\end{equation}

\subsection*{Ansatz}

We choose a variational product ansatz of the form:
\begin{equation}
\ket{\Psi} = \ket{\psi_{\text{el}}} \otimes \bigotimes_i \ket{\alpha_i},
\end{equation}
where each $\ket{\alpha_i}$ is a phonon coherent state:
\begin{equation}
\ket{\alpha_i} = e^{-|\alpha_i|^2/2} \sum_{n=0}^\infty \frac{\alpha_i^n}{\sqrt{n!}} \ket{n}.
\end{equation}

\subsection*{Expectation Values}
 the total energy is:
\begin{equation}
\begin{aligned}
E &= \langle \Psi | \hat{H} | \Psi \rangle \\
  &= \langle \Psi | \hat{H}_{\mathrm{el}} | \Psi \rangle
   + \langle \Psi | \sum_i \omega_0\, \hat{b}_i^\dagger \hat{b}_i | \Psi \rangle \\
  &\quad + \langle \Psi | g \sum_i \hat{n}_i \bigl( \hat{b}_i^\dagger + \hat{b}_i \bigr) | \Psi \rangle .
\end{aligned}
\end{equation}

\begin{equation*}
\begin{aligned}
\langle \Psi | \hat{H}_{\mathrm{el}} | \Psi \rangle
&= \bigl( \langle \psi_{\mathrm{el}} | \otimes \bigotimes_j \langle \alpha_j | \bigr)
\, \hat{H}_{\mathrm{el}} \,
\bigl( | \psi_{\mathrm{el}} \rangle \otimes \bigotimes_k | \alpha_k \rangle \bigr) .
\end{aligned}
\end{equation*}

 Since $\hat{H}_{el}$ is purely an electronic Hamiltonian, it only acts on the electronic part of the wavefunction. The phonon coherent states $|\alpha_k\rangle$ are not affected by it. Therefore, we can write:
\begin{equation*}
\langle \Psi | \hat{H}_{\mathrm{el}} | \Psi \rangle
= \langle \psi_{\mathrm{el}} | \hat{H}_{\mathrm{el}} | \psi_{\mathrm{el}} \rangle\,
\bigl( \bigotimes_j \langle \alpha_j | \bigr)
\bigl( \bigotimes_k | \alpha_k \rangle \bigr) .
\end{equation*}

 The term $\bigl( \bigotimes_j \langle \alpha_j | \bigr)
\left(\bigotimes_k|\alpha_k\rangle\right)$ represents the inner product of all the phonon coherent states: $$=\langle\alpha_1|\alpha_1\rangle\langle\alpha_2|\alpha_2\rangle\dots\langle\alpha_N|\alpha_N\rangle$$ Since phonon coherent states are normalized, $\langle\alpha_j|\alpha_j\rangle=1$ for all $j$. Therefore, the product $\prod_j\langle\alpha_j|\alpha_j\rangle=1\cdot1\cdot\cdots\cdot1=1$. So, the first term simplifies to: $$\langle\Psi|\hat{H}_{el}|\Psi\rangle=\langle\psi_{el}|\hat{H}_{el}|\psi_{el}\rangle\cdot1=\langle\psi_{el}|\hat{H}_{el}|\psi_{el}\rangle$$

Let's examine the coupling term in the Hamiltonian and then its expectation value. The original coupling term in the Hamiltonian is: $$\hat{H}_{coupling}=g\sum_i\hat{n}_i(\hat{b}_i^\dagger+\hat{b}_i)$$ Now, let's consider the expectation value of this term with respect to the chosen variational product ansatz: $$|\Psi\rangle=|\psi_{el}\rangle\otimes\bigotimes_j|\alpha_j\rangle$$ So, the expectation value is:

\begin{align}
\langle \hat{H}_{\text{coupling}} \rangle
&= g \sum_i
\left\langle \psi_{\mathrm{el}} \right| \hat{n}_i \left| \psi_{\mathrm{el}} \right\rangle
\left\langle \alpha_i \right| ( \hat{b}_i^\dagger + \hat{b}_i )
\left| \alpha_i \right\rangle
\end{align}

\[
= g \sum_i
\bigl\langle \psi_{\mathrm{el}} \otimes \bigotimes_j \alpha_j \bigr|
\hat{n}_i \bigl( \hat{b}_i^\dagger + \hat{b}_i \bigr)
\bigr| \psi_{\mathrm{el}} \otimes \bigotimes_k \alpha_k \bigr\rangle .
\]

Now, consider a single term in the sum, for a specific site $i$: 


\[
\langle \psi_{\mathrm{el}} \otimes \bigotimes_j \alpha_j |
\hat{n}_i \bigl( \hat{b}_i^\dagger + \hat{b}_i \bigr)
| \psi_{\mathrm{el}} \otimes \bigotimes_k \alpha_k \rangle
\]

Since the ansatz separates the electronic and phononic parts, and the phonon coherent states for different sites are independent, we can separate the expectation values: $$=\langle\psi_{el}|\hat{n}_i|\psi_{el}\rangle\cdot\langle\alpha_i|(\hat{b}_i^\dagger+\hat{b}_i)|\alpha_i\rangle\cdot\prod_{j\neq i}\langle\alpha_j|\hat{I}|\alpha_j\rangle$$ The product $\prod_{j\neq i}\langle\alpha_j|\hat{I}|\alpha_j\rangle$ simplifies to 1 because coherent states are normalized ($\langle\alpha_j|\alpha_j\rangle=1$). So, for each site $i$, the term becomes: $$\langle\psi_{el}|\hat{n}_i|\psi_{el}\rangle\cdot\langle\alpha_i|(\hat{b}_i^\dagger+\hat{b}_i)|\alpha_i\rangle$$ We know that: $$\langle\alpha_i|(\hat{b}_i^\dagger+\hat{b}_i)|\alpha_i\rangle=2\text{Re}(\alpha_i)$$ And the term $\langle\psi_{el}|\hat{n}_i|\psi_{el}\rangle$ is precisely the expectation value of the number operator $\hat{n}_i$ in the electronic ground state . This is denoted as $\langle \hat{n}_i\rangle$. Therefore, substituting these back into the sum, the coupling term in the total energy expression becomes: $$g\sum_i\langle \hat{n}_i\rangle(2\text{Re}(\alpha_i))$$ 

We start with the free phonon part of the Hamiltonian and the variational ansatz for the phononic part of the system. The free phonon Hamiltonian is: $$\hat{H}_{ph} = \sum_i \omega_0 \hat{b}_i^\dagger \hat{b}_i$$ The variational ansatz for the phononic part is a product of coherent states: $$|\Psi_{ph}\rangle = \bigotimes_j |\alpha_j\rangle$$ The expectation value of the phonon Hamiltonian with respect to this ansatz is:

\[
\langle \Psi_{\mathrm{ph}} | \hat{H}_{\mathrm{ph}} | \Psi_{\mathrm{ph}} \rangle
= \bigl\langle \bigotimes_k \alpha_k \bigr|
\sum_i \omega_0\, \hat{b}_i^\dagger \hat{b}_i
\bigr| \bigotimes_j \alpha_j \bigr\rangle .
\]

Due to the product nature of the coherent state ansatz, the expectation value of the operator acting on site $i$ simplifies. The terms for all other sites ($j \neq i$) will simply be $\langle \alpha_j | \alpha_j \rangle = 1$ due to the normalization of the coherent states. This leaves us with just the term for site $i$: $$ \sum_i \omega_0 \langle \alpha_i | \hat{b}_i^\dagger \hat{b}_i | \alpha_i \rangle$$ Now we use the property of coherent states, where $|\alpha_i\rangle$ is an eigenstate of the annihilation operator, $b_i |\alpha_i\rangle = \alpha_i |\alpha_i\rangle$, and its conjugate, $\langle \alpha_i | b_i^\dagger = \alpha_i^* \langle \alpha_i |$. Substituting this: 
  $$\sum_i \omega_0 \alpha_i^* \alpha_i \langle \alpha_i | \alpha_i \rangle$$ Since the coherent states are normalized, $\langle \alpha_i | \alpha_i \rangle = 1$. And by definition, $\alpha_i^* \alpha_i = |\alpha_i|^2$. This simplifies to: $$\sum_i \omega_0 |\alpha_i|^2$$

therefore, the total energy becomes:
\begin{equation}
E[\{\alpha_i\}] =
\bra{\psi_{\text{el}}} \hat{H}_{\text{el}} \ket{\psi_{\text{el}}}
+ \sum_i \omega_0 |\alpha_i|^2
+ \sum_i g(\alpha_i^*+\alpha_i)\langle \hat{n}_i \rangle.
\end{equation}

\subsection*{Energy Minimization}

We minimize $E[\{\alpha_i\}]$ with respect to $\alpha_i^*$:
\begin{align}
\frac{\partial E}{\partial \alpha_i^*} = \omega_0 \alpha_i + g \langle \hat{n}_i \rangle = 0 
~~\Rightarrow \quad \boxed{
\alpha_i^\star = -\frac{g}{\omega_0} \langle \hat{n}_i \rangle
}
\end{align}

\subsection*{Phonon Contribution to Energy}

Substituting $\alpha_i^\star$ into the energy yields the effective phonon-induced term:

$$ E_{\text{ph}} = \omega_0 \left|-\frac{g}{\omega_0} \langle\hat{n_i}\rangle\right|^2 + g\langle\hat{n_i}\rangle\left(-\frac{g}{\omega_0}\langle\hat{n_i}\rangle + -\frac{g}{\omega_0}\langle\hat{n_i}\rangle\right) $$


\begin{equation}
E_{\text{ph}} = -\sum_i \frac{g^2}{\omega_0} \langle \hat{n}_i \rangle^2
\end{equation}

The mean-field decoupling of the electron--phonon interaction
introduces the correction term
\begin{equation}
E_{\mathrm{ph}}^{(2)}
= -g \sum_{i,\sigma}
\langle \hat n_{i\sigma} \rangle
\langle \hat b_i + \hat b_i^\dagger \rangle .
\end{equation}

Assuming a product of local phonon coherent states
$\lvert \{\alpha_i\} \rangle$, we use
\begin{equation}
\langle \hat b_i + \hat b_i^\dagger \rangle
= \alpha_i + \alpha_i^\ast .
\end{equation}
Minimization of the total energy with respect to $\alpha_i^\ast$ yields the
stationary displacement
\begin{equation}
\alpha_i^\ast
= -\frac{g}{\omega_0} \langle \hat n_i \rangle ,
\end{equation}
which is real at the minimum.

Substituting this result into $E_{\mathrm{ph}}^{(2)}$ gives
\begin{equation}
E_{\mathrm{ph}}^{(2)}
= 2 \sum_i \frac{g^2}{\omega_0}
\langle \hat n_i \rangle^2 .
\end{equation}

Combining the two phonon contributions, we obtain
\begin{align}
E_{\mathrm{ph}}^{\mathrm{tot}}
&= E_{\mathrm{ph}}^{(1)} + E_{\mathrm{ph}}^{(2)} \\
&= -\sum_i \frac{g^2}{\omega_0} \langle \hat n_i \rangle^2
+ 2 \sum_i \frac{g^2}{\omega_0} \langle \hat n_i \rangle^2 \\
&= \sum_i \frac{g^2}{\omega_0} \langle \hat n_i \rangle^2 .
\end{align}

Finally, using $\alpha_i = -(g/\omega_0)\langle \hat n_i \rangle$, the total
phonon energy can be written compactly as
\begin{equation}
E_{\mathrm{ph}}^{\mathrm{tot}}
= \omega_0 \sum_i \lvert \alpha_i \rvert^2 .
\end{equation}

\section*{REFERENCES}
\bibliographystyle{aipnum4-2}
\bibliography{refs}

@inproceedings{access,
author = {Boerner, Timothy J. and Deems, Stephen and Furlani, Thomas R. and Knuth, Shelley L. and Towns, John},
title = {ACCESS: Advancing Innovation: NSF’s Advanced Cyberinfrastructure Coordination Ecosystem: Services \& Support},
year = {2023},
isbn = {9781450399852},
publisher = {Association for Computing Machinery},
address = {New York, NY, USA},
url = {https://doi.org/10.1145/3569951.3597559},
doi = {10.1145/3569951.3597559},
booktitle = {Practice and Experience in Advanced Research Computing 2023: Computing for the Common Good},
pages = {173–176},
numpages = {4},
location = {Portland, OR, USA},
series = {PEARC '23}
}

@article{Berger1995,
  author    = {E. Berger and P. Val\'{a}\v{s}ek and W. von der Linden},
  title     = {Two-Dimensional Hubbard–Holstein Model},
  journal   = {Physical Review B},
  year      = {1995},
  volume    = {52},
  
  pages     = {4806},
  doi       = {10.1103/PhysRevB.52.4806},
  url       = {https://doi.org/10.1103/PhysRevB.52.4806}
}

@article{Liu2023JCTC,
  author  = {Liu, Yuan and Meitei, Oinam R. and Chin, Zachary E. and Dutt, Arkopal and Tao, Max and Chuang, Isaac L. and Van Voorhis, Troy},
  title   = {Bootstrap Embedding on a Quantum Computer},
  journal = {J. Chem. Theory Comput.},
  volume  = {19},
  number  = {8},
  pages   = {2230--2247},
  year    = {2023},
  doi     = {10.1021/acs.jctc.3c00012},
  url     = {https://pubs.acs.org/doi/10.1021/acs.jctc.3c00012}
}

@misc{Bierman2025ChemRxiv,
  author       = {Bierman, Joel and Liu, Yuan},
  title        = {Towards Utility-Scale Electronic Structure with Sample-Based Quantum Bootstrap Embedding},
  year         = {2025},
  howpublished = {\url{https://chemrxiv.org/engage/chemrxiv/article-details/68c8d7c69008f1a467ce73f7}},
  note         = {ChemRxiv preprint, accessed December 30, 2025}
}

@article{Meitei2023PeriodicBE,
  author  = {Meitei, Oinam Romesh and Van Voorhis, Troy},
  title   = {Periodic Bootstrap Embedding},
  journal = {J. Chem. Theory Comput.},
  volume  = {19},
  number  = {11},
  pages   = {3123--3130},
  year    = {2023},
  doi     = {10.1021/acs.jctc.3c00069},
  url     = {https://pubs.acs.org/doi/10.1021/acs.jctc.3c00069}
}

@article{Ye2019JCTC,
  author  = {Ye, Hong-Zhou and Ricke, Nathan D. and Tran, Henry K. and Van Voorhis, Troy},
  title   = {Bootstrap Embedding for Molecules},
  journal = {J. Chem. Theory Comput.},
  volume  = {15},
  number  = {8},
  pages   = {4497--4506},
  year    = {2019},
  doi     = {10.1021/acs.jctc.9b00529},
  url     = {https://pubs.acs.org/doi/10.1021/acs.jctc.9b00529}
}

@article{Tezuka2007,
  author  = {Tezuka, M. and Arita, R. and Aoki, H.},
  title   = {Phase diagram for the one-dimensional Hubbard–Holstein model: A density-matrix renormalization group study},
  journal = {Phys. Rev. B},
  volume  = {76},
  issue   = {15},
  pages   = {155114},
  year    = {2007},
  doi     = {https://doi.org/10.1103/PhysRevB.76.155114}
}

@article{Knizia2012,
  author    = {Gerald Knizia and Garnet Kin-Lic Chan},
  title     = {Density Matrix Embedding: A Simple Alternative to Dynamical Mean-Field Theory},
  journal   = {Physical Review Letters},
  year      = {2012},
  volume    = {109},
  
  pages     = {186404},
  doi       = {10.1103/PhysRevLett.109.186404},
  url       = {https://doi.org/10.1103/PhysRevLett.109.186404}
}

@article{Wang2020,
  author    = {Yao Wang and Ilya Esterlis and Tao Shi and J. Ignacio Cirac and Eugene Demler},
  title     = {Zero‐Temperature Phases of the Two‐Dimensional Hubbard–Holstein Model: A Non‐Gaussian Exact Diagonalization Study},
  journal   = {Physical Review Research},
  year      = {2020},
  volume    = {2},
  pages     = {043258},
  doi       = {10.1103/PhysRevResearch.2.043258},
  url       = {https://doi.org/10.1103/PhysRevResearch.2.043258}
}

@article{Giustino2017,
  author    = {Feliciano Giustino},
  title     = {Electron–Phonon Interactions from First Principles},
  journal   = {Reviews of Modern Physics},
  year      = {2017},
  volume    = {89},
  
  pages     = {015003},
  doi       = {10.1103/RevModPhys.89.015003},
  url       = {https://doi.org/10.1103/RevModPhys.89.015003}
}

@article{Morosan2012,
  author    = {Emilia Morosan and Douglas Natelson and Andriy H. Nevidomskyy and Qimiao Si},
  title     = {Strongly Correlated Materials},
  journal   = {Advanced Materials},
  year      = {2012},
  volume    = {24},
  number    = {36},
  pages     = {4896-4923},
  doi       = {10.1002/adma.201202018},
  url       = {https://doi.org/10.1002/adma.201202018}
}

@article{Dobry1995,
  author    = {A. Dobry and A. Greco and S. Koval and J. Riera},
  title     = {Exact Diagonalization Study of the Two‐Dimensional t‐J–Holstein Model},
  journal   = {Physical Review B},
  year      = {1995},
  volume    = {52},
  
  pages     = {13722},
  doi       = {10.1103/PhysRevB.52.13722},
  url       = {https://doi.org/10.1103/PhysRevB.52.13722}
}

@article{Luo2025,
  author  = {Luo, Yao and Park, Jinsoo and Bernardi, Marco},
  title   = {First-principles diagrammatic Monte Carlo for electron-phonon interactions and polaron},
  journal = {Nature Physics},
  volume  = {21},
  pages   = {1275-1282},
  year    = {2025},
  doi     = {10.1038/s41567-025-02954-1}
}

@article{Foulkes2001,
  author  = {Foulkes, W. M. C. and Mitas, L. and Needs, R. J. and Rajagopal, G.},
  title   = {Quantum Monte Carlo simulations of solids},
  journal = {Rev. Mod. Phys.},
  volume  = {73},
  pages   = {33--83},
  year    = {2001},
  doi     = {10.1103/RevModPhys.73.33}
}

@article{Purwanto2004,
  author  = {Purwanto, Wirawan and Zhang, Shiwei},
  title   = {Quantum Monte Carlo method for the ground state of many-boson systems},
  journal = {Phys. Rev. E},
  volume  = {70},
  pages   = {056702},
  year    = {2004},
  doi     = {10.1103/PhysRevE.70.056702}
}

@article{Rubenstein2012,
  author  = {Rubenstein, Brenda M. and Zhang, Shiwei and Reichman, David R.},
  title   = {Finite-temperature auxiliary-field quantum Monte Carlo technique for Bose-Fermi mixtures},
  journal = {Phys. Rev. A},
  volume  = {86},
  pages   = {053606},
  year    = {2012},
  doi     = {10.1103/PhysRevA.86.053606}
}

@article{Backes2023,
  author  = {Backes, Steffen and Murakami, Yuta and Sakai, Shiro and Arita, Ryotaro},
  title   = {Dynamical mean-field theory for the Hubbard-Holstein model on a quantum device},
  journal = {Phys. Rev. B},
  volume  = {107},
  pages   = {165155},
  year    = {2023},
  doi     = {10.1103/PhysRevB.107.165155}
}

@article{Werner2007,
  author  = {Werner, Philipp and Millis, Andrew J.},
  title   = {Efficient Dynamical Mean Field Simulation of the Holstein-Hubbard Model},
  journal = {Phys. Rev. Lett.},
  volume  = {99},
  pages   = {146404},
  year    = {2007},
  doi     = {10.1103/PhysRevLett.99.146404}
}

@article{Jeon2004,
  author  = {Jeon, Gun Sang and Park, Tae-Ho and Han, Jung Hoon and Lee, Hyun C. and Choi, Han-Yong},
  title   = {Dynamical mean-field theory of the Hubbard-Holstein model at half filling: Zero temperature metal-insulator and insulator-insulator transitions},
  journal = {Phys. Rev. B},
  volume  = {70},
  pages   = {125114},
  year    = {2004},
  doi     = {10.1103/PhysRevB.70.125114}
}

@article{Macridin2018,
  author    = {Alexandru Macridin and Panagiotis Spentzouris and James Amundson and Roni Harnik},
  title     = {Electron‐Phonon Systems on a Universal Quantum Computer},
  journal   = {Physical Review Letters},
  year      = {2018},
  volume    = {121},
  pages     = {110504},
  doi       = {10.1103/PhysRevLett.121.110504},
  url       = {https://doi.org/10.1103/PhysRevLett.121.110504}
}

@article{Sawaya2020,
  author    = {Nicolas P. D. Sawaya and Tim Menke and Thi Ha Kyaw and Sonika Johri and Al\'an Aspuru-Guzik and Gian Giacomo Guerreschi},
  title     = {Resource-Efficient Digital Quantum Simulation of d-Level Systems for Photonic, Vibrational, and Spin-s Hamiltonians},
  journal   = {npj Quantum Information},
  year      = {2020},
  volume    = {6},
  
  doi       = {10.1038/s41534-020-0278-0},
  url       = {https://doi.org/10.1038/s41534-020-0278-0}
}

@misc{Bidart2025arxiv,
  author       = {Bidart, Alan and Vaish, Prateek and Kabengele, Tilas and Pang, Yaoqi and Liu, Yuan and Rubenstein, Brenda M.},
  title        = {Quantum Computing Beyond Ground State Electronic Structure: A Review of Progress Toward Quantum Chemistry Out of the Ground State},
  year         = {2025},
  howpublished = {\url{https://arxiv.org/pdf/2509.19709.pdf}},
  note         = {arXiv:2509.19709], accessed December 30, 2025}
}

@misc{Liu2024arxiv,
  author       = {Liu, Yuan and Singh, Shraddha and Smith, Kevin C. and Crane, Eleanor and Martyn, John M. and Eickbusch, Alec and Schuckert, Alexander and Li, Richard D. and Sinanan-Singh, Jasmine and Soley, Micheline B. and Tsunoda, Takahiro and Chuang, Isaac L. and Wiebe, Nathan and Girvin, Steven M.},
  title        = {Hybrid Oscillator-Qubit Quantum Processors: Instruction Set Architectures, Abstract Machine Models, and Applications},
  year         = {2025},
  howpublished = {\url{https://arxiv.org/abs/2407.10381}},
  note         = {arXiv:2407.10381 [quant-ph], accessed December 30, 2025}
}

@misc{Crane2024arxiv2409,
  author       = {Crane, Eleanor and Smith, Kevin C. and Tomesh, Teague and Eickbusch, Alec and Martyn, John M. and Kühn, Stefan and Funcke, Lena and DeMarco, Michael Austin and Chuang, Isaac L. and Wiebe, Nathan and Schuckert, Alexander and Girvin, Steven M.},
  title        = {Hybrid Oscillator-Qubit Quantum Processors: Simulating Fermions, Bosons, and Gauge Fields},
  year         = {2024},
  howpublished = {\url{https://arxiv.org/abs/2409.03747}},
  note         = {arXiv:2409.03747 [quant-ph], accessed December 30, 2025}
}

@misc{Kemper2025arxiv2511,
  author       = {Kemper, A. F. and Alvertis, Antonios and Asaduzzaman, Muhammad and Bakalov, Bojko N. and Baron, Dror and Bierman, Joel and Burgstahler, Blake and Chundury, Srikar and Das, Elin Ranjan and Furches, Jim and Guo, Fucheng and Jha, Raghav G. and Klymko, Katherine and Kushwaha, Arvin and Li, Ang and Majumdar, Aishwarya and Ortiz Marrero, Carlos and Mohapatra, Shubdeep and Mori, Christopher and Mueller, Frank and Popovici, Doru Thom and Stavenger, Tim and Tirfe, Mastawal and Tubman, Norm M. and Zheng, Muqing and Zhou, Huiyang and Liu, Yuan},
  title        = {Hybrid continuous-discrete-variable quantum computing: a guide to utility},
  year         = {2025},
  howpublished = {\url{https://arxiv.org/abs/2511.13882}},
  note         = {arXiv:2511.13882 [quant-ph], accessed December 30, 2025}
}

@article{Jones2020JACS,
  author  = {Jones, Leighton O. and Mosquera, Mart\'{i}n A. and Schatz, George C. and Ratner, Mark A.},
  title   = {Embedding Methods for Quantum Chemistry: Applications from Materials to Life Sciences},
  journal = {J. Am. Chem. Soc.},
  volume  = {142},
  pages   = {3281--3295},
  year    = {2020},
  doi     = {10.1021/jacs.9b10780},
  url     = {https://pubs.acs.org/doi/10.1021/jacs.9b10780}
}

@article{Rossmannek2023JPCL,
  author  = {Rossmannek, Max and Pavo\v{s}evi\'{c}, Fabijan and Rubio, Angel and Tavernelli, Ivano},
  title   = {Quantum Embedding Method for the Simulation of Strongly Correlated Systems on Quantum Computers},
  journal = {J. Phys. Chem. Lett.},
  volume  = {14},
  number  = {14},
  pages   = {3491--3497},
  year    = {2023},
  doi     = {10.1021/acs.jpclett.3c00330},
  url     = {https://pubs.acs.org/doi/10.1021/acs.jpclett.3c00330}
}

@article{SunChan2016AccChemRes,
  author  = {Sun, Qiming and Chan, Garnet Kin-Lic},
  title   = {Quantum Embedding Theories},
  journal = {Accounts of Chemical Research},
  volume  = {49},
  number  = {12},
  pages   = {2705--2712},
  year    = {2016},
  doi     = {10.1021/acs.accounts.6b00356},
  url     = {https://pubs.acs.org/doi/10.1021/acs.accounts.6b00356}
}

@article{Sandhoefer2016,
  title={Density matrix embedding theory for interacting electron-phonon systems},
  author={Barbara Sandhoefer and Garnet Kin-Lic Chan},
  journal={Physical Review B},
  volume={94},
  
  pages={085115},
  year={2016},
  publisher={APS},
  url={https://doi.org/10.1103/PhysRevB.94.085115}
}

@article{McArdle2020,
  author    = {S. McArdle and others},
  title     = {Quantum Computational Chemistry},
  journal   = {Reviews of Modern Physics},
  year      = {2020},
  volume    = {92},
  pages     = {015003},
  doi       = {10.1103/RevModPhys.92.015003},
  url       = {https://doi.org/10.1103/RevModPhys.92.015003}
}

@article{Jansen2020,
  author  = {Jansen, David and Bon\v{c}a, Janez and Heidrich-Meisner, Fabian},
  title   = {Finite-temperature density-matrix renormalization group method for electron-phonon systems: Thermodynamics and Holstein-polaron spectral functions},
  journal = {Phys. Rev. B},
  volume  = {102},
  pages   = {165155},
  year    = {2020},
  doi     = {https://doi.org/10.1103/PhysRevB.102.165155}
}

@article{quemb,
  author  = {Minsik Cho and Oinam Romesh Meitei and Leah P. Weisburn and Oskar Weser and Shaun Weatherly and Alexandra Alexiu and Rebecca Hanscam and Henry K. Tran and Hong-Zhou Ye and Matthew Welborn and Nathan Ricke and Takashi Tsuchimochi and Aleksandr Trofimov and Temujin Orkhon and Noah Whelpley and Carina Luo and Troy Van Voorhis},
  title   = {QuEmb: A Toolbox for Bootstrap Embedding Calculations of Molecular and Periodic Systems},
  journal = {Journal of Physical Chemistry A},
  year    = {2025},
  volume  = {129},
  number  = {28},
  
  url={https://doi.org/10.1021/acs.jpca.5c02983}
  
}

@misc{github_quemb,
  author       = {Meitei, Oinam Romesh and contributors},
  title        = {QuEmb: A Toolbox for Bootstrap Embedding Calculations of Molecular and Periodic Systems},
  year         = {2025},
  howpublished = {\url{https://github.com/oimeitei/quemb}},
  note         = {GitHub repository, accessed December 30, 2025}
}

@misc{block2,
  author       = {Block2 Developers},
  title        = {Custom Hamiltonians — Block2 Documentation},
  howpublished = {\url{https://block2.readthedocs.io/en/latest/tutorial/custom-hamiltonians.html}},
  year         = {2025},
  note         = {Accessed December 30, 2025}
}

@article{ Tezuka2005,
  title={A DMRG study of correlation functions in the Holstein–Hubbard model},
  author={ Tezuka,Masaki and Arita, Ryotaro  and  Aoki,Hideo},
  journal={Physica B: Condensed Matter},
  volume={359-361},
  
  pages={708-710},
  year={2005},
url={https://doi.org/10.1016/j.physb.2005.01.201},
  publisher={Elsevier}
}

@article{Clay2005,
  author    = {R. T. Clay and R. P. Hardikar},
  title     = {Intermediate Phase of the One‐Dimensional Half‐Filled Hubbard–Holstein Model},
  journal   = {Physical Review Letters},
  year      = {2005},
  volume    = {95},
  
  pages     = {096401},
  doi       = {10.1103/PhysRevLett.95.096401},
  url       = {https://doi.org/10.1103/PhysRevLett.95.096401}
}

@article{Costa2020,
  author    = {Natanael C. Costa and Kazuhiro Seki and Seiji Yunoki and Sandro Sorella },
  title     = {Phase Diagram of the Two-Dimensional Hubbard–Holstein Model},
  journal   = {Communications Physics},
  year      = {2020},
  volume    = {3},
  pages     = {80},
  doi       = {10.1038/s42005-020-0342-2},
  url       = {https://doi.org/10.1038/s42005-020-0342-2}
}

@article{Hubbard1963,
  author    = {J. Hubbard},
  title     = {Electron Correlations in Narrow Energy Bands},
  journal   = {Proceedings A},
  year      = {1963},
  volume    = {276},
  
  pages     = {238-257},
  doi       = {10.1098/rspa.1963.0204},
  url       = {https://doi.org/10.1098/rspa.1963.0204}
}

@article{Hardikar2007,
  author    = {R. P. Hardikar and R. T. Clay},
  title     = {Phase Diagram of the One‐Dimensional Hubbard–Holstein Model at Half and Quarter Filling},
  journal   = {Physical Review B},
  year      = {2007},
  volume    = {75},
  
  pages     = {245103},
  doi       = {10.1103/PhysRevB.75.245103},
  url       = {https://doi.org/10.1103/PhysRevB.75.245103}
}

@article{Dagotto2008,
  author    = {E. Dagotto and Y. Tokura},
  title     = {Strongly Correlated Electronic Materials: Present and Future},
  journal   = {MRS Bulletin},
  year      = {2008},
  volume    = {33},
  
  
  doi       = {10.1557/mrs2008.223},
  url       = {https://doi.org/10.1557/mrs2008.223}
}

@article{Jeckelmann1999,
  title={Metal-insulator transition in the one-dimensional Hubbard-Holstein model},
  author={Eric Jeckelmann and Chunli Zhang and Steven R. White},
  journal={Physical Review B},
  volume={60},
  number={11},
  pages={7950},
  year={1999},
  publisher={APS},
  url={https://doi.org/10.1103/PhysRevB.60.7950}
}

@article{QWang1999,
  title={Quantum lattice fluctuations in a model electron–phonon system},
  author={Q. Wang and H. Zheng},
  journal={Physics Letters A},
  volume={260},
  pages={99-107},
  year={1999},
  publisher={Elsevier},
  url={https://doi.org/10.1016/S0375-9601(99)00489-2}
}

@article{Holstein1959,
  author  = {Holstein, T.},
  title   = {Studies of polaron motion: Part I. The molecular-crystal model},
  journal = {Annals of Physics},
  volume  = {8},
  pages   = {325-342},
  year    = {1959},
  doi     = {10.1016/0003-4916(59)90002-8},
  url     = {https://www.sciencedirect.com/science/article/pii/0003491659900028}
}

@article{Andersen2015,
  author  = {Andersen, Ulrik L. and Neergaard-Nielsen, Jonas S. and van Loock, Peter and Furusawa, Akira},
  title   = {Hybrid discrete- and continuous-variable quantum information},
  journal = {Nature Physics},
  volume  = {11},
  pages   = {713--719},
  year    = {2015},
  doi     = {10.1038/nphys3410},
  url     = {https://www.nature.com/articles/nphys3410}
}

@article{Liu2023,
  title={Bootstrap embedding on a quantum computer},
  author={Liu, Yuan and Meitei, Oinam R and Chin, Zachary E and Dutt, Arkopal and Tao, Max and Chuang, Isaac L and Van Voorhis, Troy},
  journal={arXiv preprint arXiv:2301.01457},
  year={2023}
}

@article{Jankovic2022,
  author    = {Veljko Jankovi\'{c} and Nenad Vukmirovi\'{c}},
  title     = {Spectral and Thermodynamic Properties of the Holstein Polaron: Hierarchical Equations of Motion Approach},
  journal   = {Physical Review B},
  year      = {2022},
  volume    = {105},
  pages     = {054311},
  doi       = {10.1103/PhysRevB.105.054311},
  url       = {https://doi.org/10.1103/PhysRevB.105.054311}
}

@article{Reinhard2019,
  author  = {Reinhard, Teresa E. and Mordovina, Uliana and Hubig, Claudius and Kretchmer, Joshua S. and Schollw\"ock, Ulrich and Appel, Heiko and Sentef, Michael A. and Rubio, Angel},
  title   = {Density-Matrix Embedding Theory Study of the One-Dimensional Hubbard-Holstein Model},
  journal = {J. Chem. Theory Comput.},
  volume  = {15},
  number  = {4},
  pages   = {2221-2232},
  year    = {2019},
  doi     = {10.1021/acs.jctc.8b01116},
  url     = {https://pubs.acs.org/doi/10.1021/acs.jctc.8b01116}
}

@article{Li2024PRL,
  author  = {Li, Jiachen and Zhu, Tianyu},
  title   = {Interacting-Bath Dynamical Embedding for Capturing Nonlocal Electron Correlation in Solids},
  journal = {Phys. Rev. Lett.},
  volume  = {133},
  pages   = {216402},
  year    = {2024},
  doi     = {10.1103/PhysRevLett.133.216402},
  url     = {https://journals.aps.org/prl/abstract/10.1103/PhysRevLett.133.216402}
}

@misc{Joven2025arxiv,
  author       = {Joven, Kevin J. and Das, Elin Ranjan and Bierman, Joel and Majumdar, Aishwarya and Hakimi Heris, Masoud and Liu, Yuan},
  title        = {Scalable Quantum Computational Science: A Perspective from Block-Encodings and Polynomial Transformations},
  year         = {2025},
  howpublished = {\url{https://arxiv.org/abs/2511.16738}},
  doi          = {10.48550/arXiv.2511.16738},
  note         = {arXiv:2511.16738 [quant-ph], accessed December 30, 2025}
}

@misc{fbBE_2025,
  author       = {Shariful Islam, Joel Bierman, Yuan Liu},
  title        = {fb-{BE}},
  howpublished = {GitHub repository},
  year         = {2025},
  url          = {https://github.com/Shariful38/fb-BE}
}

\end{document}